\documentclass{article}

\usepackage{PRIMEarxiv}

\usepackage[utf8]{inputenc} % allow utf-8 input
\usepackage[T1]{fontenc}    % use 8-bit T1 fonts
\usepackage{hyperref}       % hyperlinks
\usepackage{url}            % simple URL typesetting
\usepackage{booktabs}       % professional-quality tables
\usepackage{amsfonts}       % blackboard math symbols
\usepackage{nicefrac}       % compact symbols for 1/2, etc.
\usepackage{microtype}      % microtypography
\usepackage{lipsum}
\usepackage{fancyhdr}       % header
\usepackage{graphicx}       % graphics
\graphicspath{{media/}}     % organize your images and other figures under media/ folder
\usepackage{tabularx}
\usepackage{multirow}
\usepackage{makecell}
\usepackage{caption}
\usepackage{subcaption}

\title{STEC: See-Through Transformer-based Encoder for CTR Prediction
}

\author{
 Serdarcan Dilbaz \\
  AI Enablement\\
  Huawei Türkiye R\&D Center\\
  Istanbul, Turkey \\
  \texttt{serdarcan.dilbaz@huawei.com} \\
   \And
 Hasan Saribas \\
  AI Enablement\\
  Huawei Türkiye R\&D Center\\
  Istanbul, Turkey \\
  \texttt{hasan.saribas1@huawei.com}
}

\begin{document}
\maketitle

\begin{abstract}
Click-Through Rate (CTR) prediction holds a pivotal place in online advertising and recommender systems since CTR prediction performance directly influences the overall satisfaction of the users and the revenue generated by companies. Even so, CTR prediction is still an active area of research since it involves accurately modelling the preferences of users based on sparse and high-dimensional features where the higher-order interactions of multiple features can lead to different outcomes.

Most CTR prediction models have relied on a single fusion and interaction learning strategy. The few CTR prediction models that have utilized multiple interaction modelling strategies have treated each interaction to be self-contained. In this paper, we propose a novel model named STEC that reaps the benefits of multiple interaction learning approaches in a single unified architecture. Additionally, our model introduces residual connections from different orders of interactions which boosts the performance by allowing lower level interactions to directly affect the predictions. Through extensive experiments on four real-world datasets, we demonstrate that STEC outperforms existing state-of-the-art approaches for CTR prediction thanks to its greater expressive capabilities.
\end{abstract}

\section{Introduction}

\begin{figure}[th]
\centering
\includegraphics[width=0.4\columnwidth]{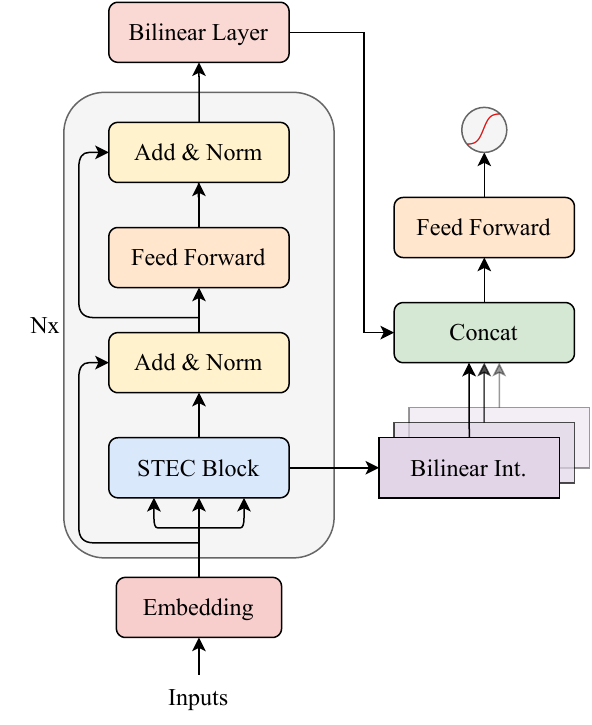}
\caption{The overall STEC architecture uses $N$ stacked STEC blocks and fuses $N+1$ group bilinear interactions from different levels to form a single CTR prediction.}
\label{fig:architecture}
\end{figure}

Click-through rate (CTR) prediction is a crucial task in the fields of online advertising and recommender systems, as it aims to forecast the probability that users will engage with displayed ads or content on online platforms \cite{wu2019session, yu2020tagnn}. Despite its extensive application across a variety of domains, including web search, recommendation systems, and online advertising, CTR prediction remains a challenging task. This is due to numerous factors such as sparse and high-dimensional feature spaces, the need to capture complex higher-order feature interactions, imbalanced data distributions, temporal dynamics, and shifts in user behavior. Accurate CTR prediction is of significant importance not only for advertisers but also for online platforms seeking to optimize user engagement and revenue generation \cite{autoint2019}. By accurately measuring the likelihood of user engagement with displayed ads, CTR prediction models enable advertisers to effectively allocate resources and adapt their campaigns to target specific user segments. Simultaneously, by serving ads that take individual preferences into account, platforms can improve user experiences and increase the value of the overall ecosystem \cite{fignn2019, liu2015convolutional}.

Historically, traditional methodologies for click-through rate (CTR) prediction have relied on statistical techniques and manually designed features. Despite the complex, non-linear interactions present in user behavior and contextual variables, these methods have achieved some measure of success \cite{rendle2010factorization, mcmahan2013ad}. One such technique is Factorization Machines (FM), which have demonstrated effectiveness in capturing second-order feature interactions by computing the inner product of feature embeddings within each interaction \cite{rendle2010factorization}. This approach has proven successful in addressing the challenges posed by CTR prediction.

Recent advancements in deep learning models have led to significant achievements in various domains, including computer vision (CV) \cite{krizhevsky2012imagenet, simonyan2014very, szegedy2015going} and natural language processing (NLP) \cite{chai2019deep, goldberg2022neural}. Methods developed specifically for these domains have also contributed to solving problems in other fields. In the area of recommendation, numerous deep learning-based methods have primarily focused on more effectively capturing higher-order feature interactions \cite{guo2017deepfm, cheng2016wide, yu2020tagnn, lian2018xdeepfm}. After achieving notable results using these methods, approaches such as FiBiNET \cite{fibinet2019} and MaskNet \cite{masknet2021} have attempted to better determine the significance of embeddings using attention mechanisms such as the Squeeze-Excitation network (SENET) \cite{hu2018squeeze}. Following the successful application of attentional mechanisms in computer vision and natural language processing, attention-based methods such as AFM, AutoInt, and DESTINE have been developed for CTR prediction with promising results. However, methods like AutoInt and DESTINE have a high parameter count, which has made their applicability challenging.

One of the primary challenges in click-through rate (CTR) prediction is the effective modeling of feature interactions \cite{guo2017deepfm, masknet2021, yu2020tagnn}. FM \cite{rendle2010factorization} has proposed a successful method for capturing second-order feature interactions through the use of inner product calculations. DeepFM \cite{guo2017deepfm} combines an FM module with a deep neural network to achieve improved performance in the CTR domain. To model high-order feature interactions, xDeepFM \cite{lian2018xdeepfm} introduced the Compressed Interaction Network (CIN) layer, which utilizes outer-product calculations. Other models, such as HOFM \cite{blondel2016higher} and AFN \cite{aoanet2020}, have also been used to extract high-order feature interactions. However, these models are not efficient in terms of time complexity. To address this issue, DeepIM \cite{yu2020tagnn} proposed the use of an IM layer instead of an FM layer to compute higher-order interactions using Newton’s identities for increased efficiency.

In this paper, we present a rigorous proof demonstrating that bilinear interactions can be derived from the commonly used scaled dot-product attention. Building on this insight, we introduce a novel model, STEC, that integrates multiple interaction learning approaches into a single, unified architecture by leveraging the bilinear interactions present in attention computations. Furthermore, our model incorporates residual connections among various orders of interactions, enhancing performance by enabling lower-level interactions to directly impact predictions. Extensive experiments on four real-world datasets and industrial offline and online evaluations demonstrate that STEC surpasses existing state-of-the-art methods for CTR prediction due to its superior expressive capabilities. We also show that STEC can outperform state-of-the-art attention-based models while being more lightweight. An ablation study on the STEC architecture confirms the effectiveness of the novel concepts introduced.
Our main contributions can be listed as follows:
\begin{itemize}
  \item To the best of our knowledge, we are the first to expose and leverage the bilinear interaction hidden inside the scaled dot-product attention calculations to model higher-order feature interactions.
  \item As opposed to the existing approaches, STEC also introduces direct connections between different interaction levels and the output which increases the performance compared to the traditional architecture where lower order interactions affect the prediction by only influencing higher order interactions.
  \item Based on the proposed modified attention mechanism, we proposed the STEC architecture and conducted extensive experiments on public datasets and industrial offline and online evaluations which show STEC outperforms or matches state-of-the-art methods while maintaining explainability.
  %Based on the proposed altered attention mechanism, we propose the STEC architecture which outperforms existing state-of-the-art approaches. 
\end{itemize}

\section{Related Works}

\subsection{Bilinear Interaction}

As previously discussed, a variety of interaction learning approaches have been used for CTR. Nevertheless, for the purposes of this paper, bilinear interaction is of particular importance. Models utilizing bilinear interaction have achieved notable results on numerous real-world datasets owing to the effectiveness of bilinear interaction layer at learning feature importance and fine-grained feature interactions \cite{fibinet2019, fibinetpp2022}.

%\subsection{Group Interactions}

%Talk about finalmlp, autoint, destine, bst, dmin, eta, interhat, sdim, parallel masknet, aoanet gin?, sam?, ccpm (pooling seen as group interaction), dmr (pooling),  

\subsection{Attention Networks}

The attention mechanism, first introduced in deep learning as an enhancement to encoder-decoder-based neural machine translation systems in NLP \cite{attention2016}, has since been applied to a wide range of applications. Initially, attention mechanisms were applied to recurrent systems, but models that rely solely on attention have become more popular due to their parallelizability and faster training times \cite{vaswani2023attention}. Attention mechanisms and their variants have also been used in other fields, including computer vision and speech processing \cite{guo2022attention}.

In the field of sequential recommendation, where the goal is to predict a user’s next action based on their sequence of interactions, attention mechanisms have also been successfully applied since this task is already sequential in nature. Models such as DIN \cite{din2018}, DIEN \cite{dien2018}, BST \cite{bst2019}, DMIN \cite{dmin2020} have introduced novel attention mechanisms to achieve significant improvements in CTR performance.

Numerous studies have explored the use of attention as an interaction learning approach on user and item embeddings. For example, the AFM \cite{afm2017} model introduces attention to the FM mechanism to improve its performance. The HoAFM \cite{hoafm} model applies bit-wise attention to learn higher-order feature interactions. The AutoInt \cite{autoint2019} model utilizes the multi-head self-attention mechanism with residual connections to explicitly model feature interactions. The DESTINE \cite{destine2021} model employs a similar structure to AutoInt but uses disentangled attention through whitening. The InterHAt \cite{interhat2020} model employs the Transformer architecture with multi-head attention and a hierarchical structure for learning feature interactions.

\section{Our Proposed Model}

\subsection{Feature Embedding}
In CTR prediction, input instances typically comprise three types of features: user, item, and context. These features may be either sparse categorical or numerical, and input instances are represented as the sparse vector $\alpha$ where $f$ is the number of feature fields:
\begin{equation}
    \alpha=[\alpha_1,\alpha_2,...,\alpha_f ]
\end{equation}
Due to the large number of unique values in the categorical features for CTR prediction, one-hot vectors are not a practical representation for input features. Instead, these features are embedded into a lower-dimensional space to enable CTR models to learn dense representations. For a categorical feature $\alpha_i \in \mathbb{R}^{k \times 1}$, the corresponding embedding is calculated as:     
\begin{equation}
    E_i=W_e \alpha_i
\end{equation}
where $W_e \in \mathbb{R}^{d \times k}$ is the embedding matrix for $k$ unique feature values in the $d$ dimensional embedding space. Unlike the shared embedding layer used in NLP tasks, a separate embedding layer is employed for each categorical feature.
Although numerical features are more amenable to being utilized in deep learning structures, these features still need to be embedded into the same dimensional space as the categorical features to allow for straightforward calculation of feature interactions. For a numerical feature $\alpha_j \in \mathbb{R}^{1 \times 1}$ , the corresponding embedding is calculated as:
\begin{equation}
    E_j=v_j \alpha_j
\end{equation}
where $v_j \in \mathbb{R}^{d \times 1}$ is scaled by the normalized numerical feature values.
For a given instance, when both categorical and numerical features undergo feature embedding, $x \in \mathbb{R}^{d \times f}$ is formed by stacking of embedding $E \in \mathbb{R}^{d \times 1}$ from different feature fields. 

\subsection{STEC Block}

\begin{figure}[t]
     \centering
     \begin{subfigure}[b]{0.55\textwidth}
         \centering
         \includegraphics[width=0.9\columnwidth]{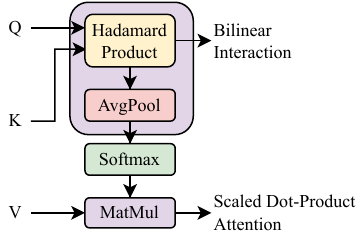}
         \caption{The STEC block builds on the Scaled Dot-Product Attention to accommodate concurrent bilinear interaction learning. The Hadamard Product and AvgPool in conjunction are equivalent to matrix multiplication.}
         \label{fig:stec_block}
     \end{subfigure}
     \hfill
     \begin{subfigure}[b]{0.35\textwidth}
         \centering
         \includegraphics[width=\textwidth]{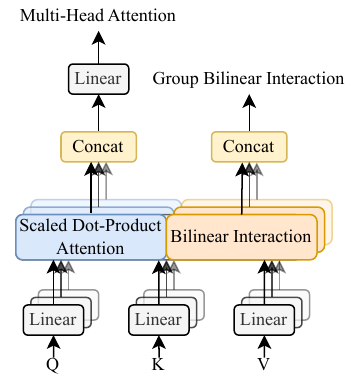}
         \caption{The STEC Block is able to run multiple attention layers and learn bilinear interactions in parallel.}
         \label{fig:multihead}
     \end{subfigure}
    \label{fig:stecinner_blocks}
\end{figure}

The core component of the STEC architecture is the STEC block, which is designed to simultaneously extract bilinear interactions by modifying the existing scaled dot-product attention formulation.
For an input $x \in \mathbb{R}^{d \times f}$ , following the scaled dot-product attention formulation, the corresponding query and key matrices are given as: 

\noindent\begin{minipage}{.5\linewidth}
\begin{equation}
  Q=W_{q}^T x
\end{equation}
\end{minipage}%
\begin{minipage}{.5\linewidth}
\begin{equation}
  K=W_{k}^T x
\end{equation}
\end{minipage}

where the learned projection matrices for the query and key are denoted as $W_{q}, W_{k} \in \mathbb{R}^{d \times d}$ respectively.
The scaled dot-product attention $\mathrm{A}$ is given by:
\begin{equation}\mathrm{A}=\textup{softmax}\left ( \frac{K^T Q}{\sqrt{d}}\right ) \end{equation}
The learnable projection matrices for the query and the key can be reduced to a single learnable parameter for simplicity and we introduce $P$ to denote the inner product of the query and the key.
\begin{equation} P = K^T Q = x^T W_{k} {W_{q}}^T x = x_i^T W x_j\end{equation}
Each element of $P\in R^{d\times d}$ can be calculated individually as $P_{ij}$.
\begin{equation} P_{ij} = x_i^T W x_j = x_i \cdot W x_j
\end{equation}
The matrix multiplication can be expressed by using the Hadamard product which is denoted by $\odot$. The summation operation calculates the sum along the second dimension of the input matrix.
\begin{equation}
P_{ij} = x_i \cdot \sum_{m}(W_m \odot x_{mj}^T) = \sum_{m}(x_{im} \cdot W_m \odot x_{mj}^T)\end{equation} 
The term $x_i \cdot W \odot x_j^T$ in the above equation for the scaled dot-product attention corresponds to the bilinear interaction used by models like FiBiNet \cite{fibinet2019} and FiBiNet++ \cite{fibinetpp2022} models. This formulation reveals that the attention matrix in the scaled dot-product attention corresponds to the scaled version of the summation of the bilinear interaction along the second dimension. By tweaking the formulation of the existing scaled dot-product attention, the bilinear interaction can be extracted at no additional cost. Through this insight, we formulate the STEC block where the scaled dot-product attention modified to allow the intermediate bilinear interaction representation to be used for CTR prediction as seen in Figure \ref{fig:stec_block}.

Motivated by the success of multi-head attention, we also leverage the multi-head mechanism for the attention as well as the bilinear interactions as seen in Figure \ref{fig:multihead}. Grouping the bilinear interactions allows the model to jointly learn the different interaction subspaces at different positions.

\subsection{STEC Architecture}

The STEC architecture is inspired by the Transformer architecture where self-attention and position-wise feed-forward networks (FFN) are alternately stacked. Similarly, the STEC architecture interlaces multiple layers of STEC blocks and FFNs to perform CTR prediction, shown in Figure \ref{fig:architecture}. In contrast to the self-attention layer of the Transformer, the STEC layer produces two outputs simultaneously. The first output is mathematically identical to the self-attention described in the Transformer, while the second output corresponds to the bilinear interaction of the input. Bilinear interactions from STEC blocks at different levels are fused together and processed by a multilayer perceptron (MLP) to produce a final prediction.

\subsubsection{Position-wise Feed-Forward Networks}

The Transformer architecture employs FFNs between attention layers \cite{vaswani2023attention}. An FFN takes the hidden attention representation $x_h$ at layer $h$ and applies a two-layer MLP, typically using a rectified-linear activation function on the hidden layer. The FFN can be formally described as:

\begin{equation}
\textup{FFN}(x_h,W_1, W_2, b_1, b_2)=max\left (0, x_h W_1+b_1\right )W_2 +b_2
\end{equation}

\subsubsection{Final Bilinear Interaction Layer}

As seen in Figure \ref{fig:stec_block}, The STEC blocks produce two outputs: an attention output and a bilinear interaction. Due to the stacked design of the architecture, the attention output from layer $k$ is connected to the STEC block from layer $k + 1$ via a FFN layer. To utilize the attention output from the final STEC block, a standalone bilinear interaction layer is used, although it has been omitted from Figure \ref{fig:architecture} for brevity. As a result, $N$ stacked layers produce $N + 1$ bilinear interactions.

\subsubsection{Concatenation Layer}

The STEC architecture combines learned interactions from multiple orders of interaction through the concatenation operation. Since the distributions of the interactions can vary greatly, batch normalization is applied to interactions at each level before they are fused to form the representation $H$ that will be used the feed forward layer to produce the final outcome.

\begin{equation}
    H = \left [\textup{BN}\left ( p_0 \right );\textup{BN}\left ( p_1 \right );...;\textup{BN}\left ( p_{N+1} \right ) \right ]
\end{equation}

\section{Experimental Results}

In this section, we move forward to evaluate the effectiveness of our proposed approach by addressing the following research questions:

\begin{table}[h]
\centering
\begin{tabularx}{\columnwidth}{lX}
    \textbf{RQ1} & How does our proposed STEC architecture compared to the existing approaches in terms of performance? \\
    \textbf{RQ2} & Does the the STEC architecture compare to approaches that rely on attention in terms of performance as well as computational cost? \\
    \textbf{RQ3} & How does the design choices affect the performance of the proposed model? \\
    \textbf{RQ4} & What kinds of interactions does the STEC architecture learn? Is our proposed model explainable? \\
\end{tabularx}
\end{table}
%RQ4 Hyperparameter importance analysis
% attention_layers, d_model

\begin{table}[h]
\caption{Summary Statistics of the four public CTR datasets}
\label{table:datasets}
\centering
\begin{tabular}{@{}l|c|c|c@{}}
\toprule
Dataset      & \#Instance & \#Fields & \#Features (Sparse) \\ \midrule
Criteo       &  46M         & 39        & 2.1M                   \\
Avazu        & 40M         & 23        & 1.5M                   \\
MovieLens    & 2M          & 7        & 90k                   \\
Frappe       & 289k          & 10        & 5.4k                   \\ \bottomrule
\end{tabular}
\end{table}

\subsection{Experiment Setup}

\subsubsection{Datasets}

We evaluate our approach on four widely used public datasets for CTR prediction: Criteo\footnote{\url{http://labs.criteo.com/2013/12/download-terabyte-click-logs}}, Avazu\footnote{\url{https://www.kaggle.com/c/avazu-ctr-prediction/data}}, MovieLens\footnote{\url{https://grouplens.org/datasets/movielens/1m/}}, and Frappe\footnote{\url{https://www.baltrunas.info/context-aware}}. To eliminate any advantages that could be gained through random splits and preprocessing steps, we applied the same splits and preprocessing steps for all models based on \cite{aoanet2020}. 

To evaluate the performance of STEC in production, we deployed the model in our production system for advertisements, which serves millions of daily users. We evaluated the performance over one week in both the online and offline settings.  

\subsubsection{Training Objective}

The training objective for both our proposed approach and the baseline models was binary cross-entropy loss, as defined by Equation \ref{eq:binarycross} where $N$, $y_i$, $\widehat{y}_i$  represent the number of samples, the ground truth label, and the predicted label, respectively.

\begin{equation}\label{eq:binarycross}
L=-\frac{1}{N}\sum_{i=1}^{N}{y_ilog\left(\widehat{y}_i\right)}+\left(1-y_i\right)log\left(1-\widehat{y}_i\right)
\end{equation}

\subsubsection{Evaluation Metric}
In accordance with established CTR prediction conventions, we primarily evaluated the performance of our proposed approach and the baseline models using the Area Under the ROC Curve (AUC) metric. We reported our AUC results to four significant digits, as it is well established that even small increases in AUC can lead to significant improvements in CTR prediction and revenue \cite{cheng2016wide, guo2017deepfm, deepcross2017}.

Logloss is also widely used in binary classification to measure how closely the predicted probabilities match the corresponding ground truth values. We report this metric as well, as a smaller logloss indicates that the model was better able to model the probabilities.

\subsubsection{Baseline Models}

For baseline comparisons, we consider any model that has achieved competitive results on any of the four datasets. We grouped attention-based models separately to assess the effectiveness of our attention mechanism compared to existing models. Abbreviations for the models listed below were used whenever possible for brevity.

\begin{itemize}
  \item \textbf{IPNN}: Product-based Neural Networks \cite{pnn2016}
  \item \textbf{Wide\&Deep}: Wide\&Deep Learning \cite{cheng2016wide}
  \item \textbf{DCN}: Deep\&Cross Network \cite{deepcross2017}
  \item \textbf{DeepFM}: Deep Factorization Machine \cite{guo2017deepfm}
  \item \textbf{xDeepFM}: eXtreme Deep Factorization Machine \cite{lian2018xdeepfm}
  \item \textbf{FiBiNET}: Feature Importance and Bilinear feature Interaction NETwork \cite{fibinet2019}
  \item \textbf{AFN+}: Adaptive Factorization Network \cite{aoanet2020}
  \item \textbf{DeepIM}: Deep Interaction Machine \cite{deepim2020}
  \item \textbf{AOANet}: Architecture and Operation Adaptive Network \cite{aoanet2020}
  \item \textbf{DCNV2}: Improved Deep\&Cross Network \cite{dcnv2}
  \item \textbf{EDCN}: Enhanced Deep\&Cross Network \cite{edcn2021}
  \item \textbf{MaskNet}: MaskNet \cite{masknet2021}
  \item \textbf{FinalMLP}: Enhanced Two-Stream MLP Model \cite{mao2023finalmlp}
  \item \textbf{AFM}: Attentional Factorization Machines \cite{afm2017}
  \item \textbf{InterHAt}: Interpretable Click-Through Rate Prediction through
Hierarchical Attention \cite{interhat2020}
  \item \textbf{AutoInt+}: Automatic Feature Interaction Learning via Self-Attentive Neural Network \cite{autoint2019}
  \item \textbf{DESTINE}: Disentangled Self-Attentive Neural Network \cite{destine2021}
\end{itemize}

\subsubsection{Implementation Details}

The implementation of our models is in PyTorch and compatible with the open-source CTR prediction library FuxiCTR \cite{barsctr}. We use Adam optimizer with a learning rate of 0.001 where the learning rate is reduced by a factor of 10 whenever the validation logloss stops improving. The mini-batch size for all experiments is 4096. We tune the hyper-parameters such as the number of layers, dimension size, and the number of heads model for each setting with grid search. We conduct our experiments on 2 NVIDIA A40 GPUs.

\begin{table}[t]
%\centering
\caption{Performance comparisons on Avazu and Criteo datasets w.r.t. AUC and Logloss. The best results are reported as \textbf{bold} and the second best results as \underline{underlined}.}
\label{table:sota}
\resizebox{\columnwidth}{!}{
\begin{tabular}{llcccccc}
\hline
\multirow{2}{*}{Type}  & \multirow{2}{*}{Model} & \multicolumn{3}{c}{Avazu}   & \multicolumn{3}{c}{Criteo} \\ \cline{3-8}
                              &                        & AUC (\%)          & Impr.   & LogLoss            & AUC (\%)          & Impr.   & LogLoss   \\\hline \hline
\multirow{13}{*}{Base Models} & IPNN                   & 76.20             & 0.58\%  & 0.3676             & 81.34             & 0.11\%  & 0.4383     \\
                              & Wide\&Deep             & 76.31             & 0.43\%  & 0.3665             & 81.38             & 0.06\%  & 0.4380     \\
                              & DCN                    & 76.28             & 0.47\%  & 0.3664             & 81.38             & 0.06\%  & 0.4378     \\
                              & DeepFM                 & 76.48             & 0.21\%  & 0.3667             & 81.37             & 0.07\%  & 0.4381     \\ 
                              & xDeepFM                & 76.30             & 0.45\%  & 0.3671             & 81.36             & 0.09\%  & 0.4380     \\
                              & FiBiNET                & 76.22             & 0.55\%  & 0.3670             & 81.29             & 0.17\%  & 0.4388     \\
                              & AFN+                   & 76.36             & 0.37\%  & 0.3672             & 81.41             & 0.02\%  & 0.4377     \\
                              & DeepIM                 & 76.42             & 0.29\%  & 0.3670             & 81.39             & 0.05\%  & 0.4380     \\
                              & AOANet                 & 76.40             & 0.31\%  & 0.3664             & 81.41             & 0.02\%  & 0.4378     \\
                              & DCNV2                  & 76.30             & 0.45\%  & 0.3664             & 81.39             & 0.05\%  & 0.4378     \\
                              & EDCN                   & 76.26             & 0.50\%  & 0.3670             & 81.42             & 0.01\%  & \underline{0.4373}     \\
                              & MaskNet                & 76.12             & 0.68\%  & 0.3674             & 81.37             & 0.07\%  & 0.4380     \\
                              & FinalMLP               & \underline{76.58} & 0.08\%  & \underline{0.3658} & \textbf{81.48}    & -0.06\% & \textbf{0.4370}     \\ \hline
\multirow{5}{*}{\makecell{
Attention Based \\ Models}}   & AFM                    & 75.52             & 1.48\%  & 0.3705             & 80.37             & 1.32\%  & 0.4470     \\ 
                              & InterHAt               & 75.30             & 1.78\%  & 0.3721             & 80.72             & 0.88\%  & 0.4441     \\
                              & AutoInt+               & 76.39             & 0.33\%  & 0.3668             & 81.37             & 0.07\%  & 0.4379     \\
                              & DESTINE                & 76.34             & 0.39\%  & 0.3661             & 81.31             & 0.15\%  & 0.4380     \\
                              & STEC (Ours)            & \textbf{76.64}    & Base    & \textbf{0.3658}    & \underline{81.43} & Base    & 0.4379     \\ \hline \smallskip

\end{tabular}}
{\raggedright Evaluated using an unpaired t-test at a 0.01 confidence level, STEC outperformed baseline models on the Avazu dataset. \par}
\end{table}

\begin{table}[t]
%\centering
\caption{Performance comparisons on Frappe and MovieLens datasets w.r.t. AUC and Logloss. The best results are reported as \textbf{bold} and the second best results as \underline{underlined}.}
\label{table:sota2}
\resizebox{\columnwidth}{!}{
\begin{tabular}{llcccccc}
\hline
\multirow{2}{*}{Type}  & \multirow{2}{*}{Model} & \multicolumn{3}{c}{Frappe}   & \multicolumn{3}{c}{MovieLens} \\ \cline{3-8}
                              &                        & AUC (\%)          & Impr.   & LogLoss             & AUC (\%)           & Impr.   & LogLoss   \\\hline \hline
\multirow{13}{*}{Base Models} & IPNN                   & 98.30             & 0.20\%  & 0.1540              & 96.89              & 0.27\%  & 0.2095     \\
                              & Wide\&Deep             & 98.35             & 0.15\%  & 0.1490              & 96.87              & 0.29\%  & 0.2161     \\
                              & DCN                    & 98.40             & 0.10\%  & 0.1491              & 96.88              & 0.28\%  & 0.2147     \\
                              & DeepFM                 & 98.42             & 0.08\%  & 0.1482              & 96.87              & 0.29\%  & 0.2130     \\ 
                              & xDeepFM                & 98.45             & 0.05\%  & 0.1466              & 96.75              & 0.41\%  & 0.2409     \\
                              & FiBiNET                & 98.10             & 0.41\%  & 0.1941              & 95.32              & 1.92\%  & 0.2518     \\
                              & AFN+                   & 98.02             & 0.49\%  & 0.2139              & 96.37              & 0.81\%  & 0.3030     \\
                              & DeepIM                 & 98.39             & 0.11\%  & 0.1490              & 96.88              & 0.28\%  & \underline{0.2099}     \\
                              & AOANet                 & 98.44             & 0.06\%  & \textbf{0.1424}     & 96.94              & 0.22\%  & 0.2105     \\
                              & DCNV2                  & 98.45             & 0.05\%  & 0.1491              & 96.91              & 0.25\%  & 0.2147     \\
                              & EDCN                   & 98.37             & 0.13\%  & 0.1547              & 96.11              & 1.08\%  & 0.2122     \\
                              & MaskNet                & 98.32             & 0.18\%  & 0.1696              & 96.73              & 0.43\%  & 0.2364     \\
                              & FinalMLP               & \underline{98.48} & 0.02\%  & 0.1484              & \underline{97.13}  & 0.02\% & 0.2246     \\ \hline
\multirow{5}{*}{\makecell{
Attention Based \\ Models}}   & AFM                    & 96.11             & 2.49\%  & 0.2264              & 94.55              & 2.75\%  & 0.2653     \\ 
                              & InterHAt               & 97.27             & 1.26\%  & 0.1919              & 94.86              & 2.41\%  & 0.2540     \\
                              & AutoInt+               & 98.44             & 0.06\%  & 0.1490              & 96.89              & 0.27\%  & 0.2148     \\
                              & DESTINE                & 98.42             & 0.08\%  & \underline{0.1434}  & 96.86              & 0.30\%  & 0.2125     \\
                              & STEC (Ours)            & \textbf{98.50}    & Base    & 0.1477              & \textbf{97.15}  & Base    & \textbf{0.2016}     \\ \hline \smallskip

\end{tabular}}
{\raggedright Evaluated using an unpaired t-test at a 0.05 confidence level, STEC outperformed baseline models on both datasets.  \par}
\end{table}

\subsection{Quantitative Results (RQ1)}

\subsubsection{Public Datasets}

We gathered the best performing hyperparameters for all models based on BarsCTR \cite{barsctr} whenever possible. During our experiments, we noticed that the initial parameters had a significant effect on the experiment results. To minimize the effect of the random initialization, we repeated our experiments 15 times and reported averages.

Based on the results presented in Tables \ref{table:sota} and \ref{table:sota2}, we observed that some CTR models perform competitively on a select few datasets. For example, the EDCN model performs exceptionally well on the Criteo dataset but lags in performance on other datasets. Some CTR models have been specifically tuned to maximize performance on certain datasets, resulting in significant deterioration when tested on different datasets.

A select few models perform competitively across all datasets. The STEC model demonstrates its flexible expressive abilities by performing well on all four public datasets. Models such as FinalMLP leverage MLPs with up to three hidden layers to perform CTR predictions, but our STEC model achieves state-of-the-art performance even when using a single linear layer to make predictions based on bilinear interactions. Overall, the STEC model performs on par with or better than state-of-the-art models while following a single-stream structure, and its expressiveness can be adjusted based on the dataset by modifying hyperparameters such as the number of layers, hidden dimension, or MLP structure.

The proposed method achieves the best results on the Avazu, Frappe, and MovieLens datasets in terms of AUC metric by 76.64\%, 97.15\%, and 98.50\% respectively. In the case of the Criteo dataset, the proposed method obtains comparable results with FinalMLP, which involves an additional step compared to the others, as the user and item features need to specified beforehand. The proposed STEC architecture achieves equivalent results without any prior knowledge of the features.

\subsubsection{Industrial Evaluation}
We evaluate STEC in our production advertisement system, where about 400 million impressions are served daily. 
Offline improvements in terms of the AUC metric for the FinalMLP (SOTA), AutoInt (attention-based), and base (our best-performing model) models are reported in Table \ref{table:Offline}. The proposed model outperforms and especially, obtained significant AUC improvement than the SOTA FinalMLP model by 1.26\%.
% First, we validate the effectiveness of STEC in the offline setting with 5 days of data. Alongside the current best performing model (BaseModel), we also compare the performance of STEC with FinalMLP and EDCN. As shown in Table \ref{table:offline}, the AUC and the logloss are improved significantly.

\begin{table}[h]
\centering
\caption{Offline evaluation in the production.}
\label{table:Offline}
\resizebox{0.6\columnwidth}{!}{
\begin{tabular}{lccc}
\hline
             & FinalMLP & AutoInt & BaseModel \\ \hline
$\Delta$ AUC & 1.26\%   & 0.23\%  & 0.21\%  \\ \hline
\end{tabular}}
\end{table}

We conducted A/B test on an online dataset for over a week where the control group is served advertisements with the best-performing base model and the experimental group is served with the proposed STEC model. The results reported in Table \ref{table:ABTesting} in terms of commonly used online CTR metric which is the ratio of total users click to total impressions. Compared to the BaseModel, STEC achieves an average CTR increase of 0.72\% for one week. Additionally, it outperformed on a day-by-day basis for the entire week, except for one day.

\begin{table}[h]
\centering
\caption{CTR improvements in daily A/B test in production.}
\label{table:ABTesting}
\resizebox{\columnwidth}{!}{
\begin{tabular}{lcccccccc}
\hline
             & Day1             & Day2              & Day3     & Day4             & Day5             & Day6             & Day7             & Average \\ \hline
$\Delta$ CTR & \textbf{1.247\%} & \textbf{0.002\%} & -1.409\% &  \textbf{1.090\%} & \textbf{3.651\%} & \textbf{0.331\%} & \textbf{0.111\%} & \textbf{0.720\%}  \\ \hline
\end{tabular}}
\end{table}

\begin{figure}[th]
\centering
\includegraphics[width=0.9\columnwidth]{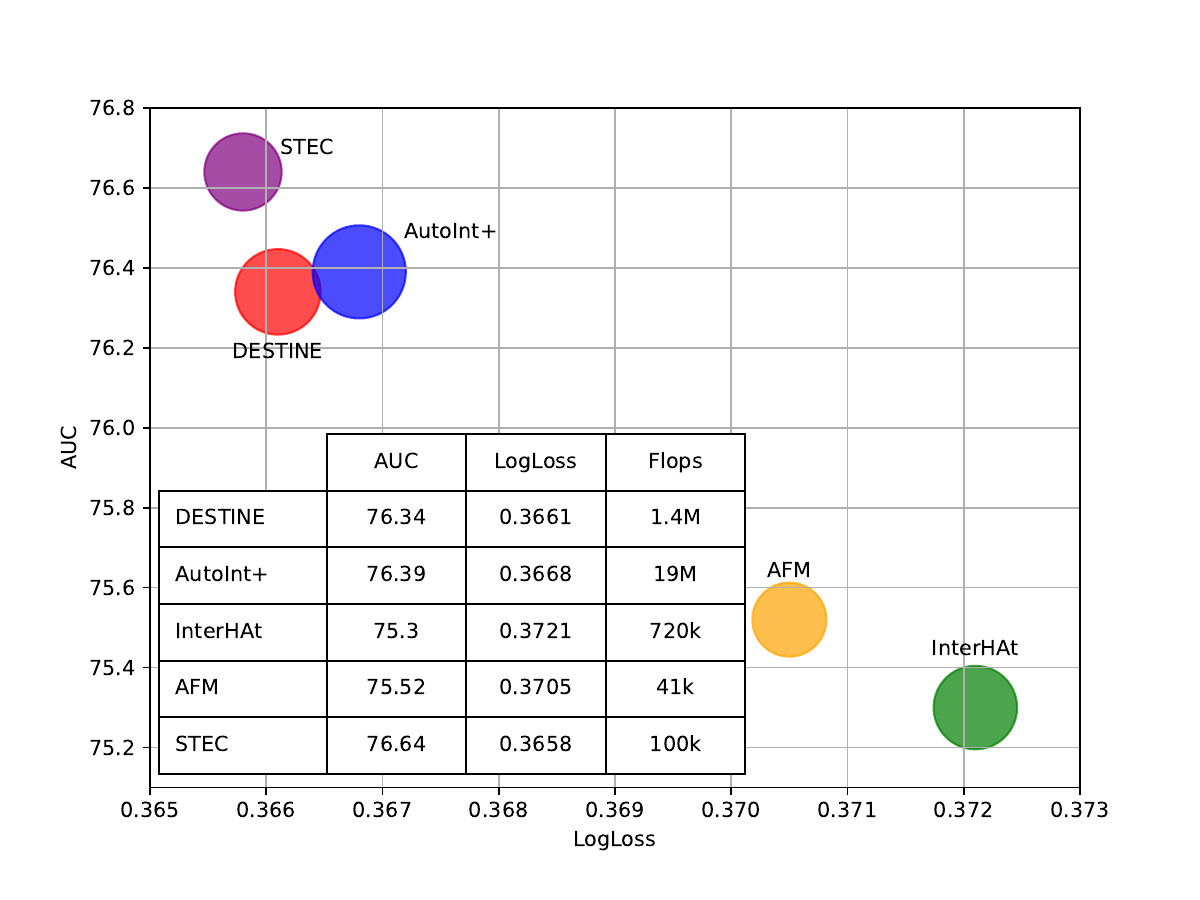}
\caption{STEC outperforms other attention-based models in terms of AUC and logloss with lower FLOPs.}
\label{fig:auc_logloss_flop_plot}
\end{figure}

\subsection{Comparison to Attention-based Models (RQ2)}

It is crucial to compare STEC with other attention-based models, as attention is central to the STEC architecture. The number of floating-point operations (FLOPs) serves as a proxy for the computational complexity of deep learning models. As shown in Figure \ref{fig:auc_logloss_flop_plot}, aside from the AFM model, the number of FLOPs for attention-based models is several orders of magnitude larger than that of the STEC model. This is due to the fact that the STEC model performs attention calculations in lower-dimensional spaces and leverages intermediate calculation steps to improve its performance.

In terms of performance, the STEC model outperforms all attention-based models across all four datasets. Our experiments show that extracting intermediate bilinear interactions improves the performance of models that rely solely on attention. The AutoInt model, which relies only on stacked multi-head attention layers, is unable to perform competitively despite its significantly higher computational cost compared to our proposed model. In the next section, we analyze the effect of using intermediate outputs instead of final outputs and discuss other novel aspects of our model.

\subsection{Analysis (RQ3)}

In this section, we conduct ablation studies on the design of the STEC architecture to investigate the impact of various design choices on its performance. We consider several variations of the STEC architecture:

\begin{itemize}
  \item \textbf{STEC\textsubscript{BL}} replaces the STEC block with a parallel multi-head attention and bilinear interaction blocks.
  \item \textbf{STEC\textsubscript{NF}} removes the FFN from the STEC\textsubscript{BL} to make the overall structure resemble the AutoInt model.
  \item \textbf{STEC\textsubscript{LO}} only uses the bilinear interaction from the final STEC block by removing the bilinear interactions from intermediate layers.
  \item \textbf{STEC\textsubscript{F}} replaces the concatenation of bilinear interactions with addition.
\end{itemize}

\begin{table}[t]
\centering
\caption{Ablation studies on Avazu, Criteo, Frappe and MovieLens datasets in terms of AUC metric. The best results are reported as \textbf{bold}.}
\label{table:ablation}
\resizebox{\columnwidth}{!}{
\begin{tabular}{l|c|c|c|c|c|c|c}
\hline
Model                  & Bilinear & Fusion    & FFN & Avazu          & Criteo         & Frappe         & MovieLens      \\ \hline
STEC\textsubscript{BL} & Exp      & Concat    & \checkmark   & 76.22          & 81.22          & 98.44          & 97.03          \\ \hline 
STEC\textsubscript{NF} & Exp      & Concat    & \textit{X}   & 76.56          & 81.25          & 98.35          & 96.89           \\ \hline
STEC\textsubscript{LO} & Imp      & Last Only & \checkmark   & 75.83          & 81.39          & 98.42          & 97.03           \\ \hline
STEC\textsubscript{F}  & Imp      & Add       & \checkmark   & 76.08          & 81.34          & 98.43          & 97.08           \\ \hline
STEC                   & Imp      & Concat    & \checkmark   & \textbf{76.64} & \textbf{81.43} & \textbf{98.50} & \textbf{97.15}  \\ \hline
\end{tabular}}
\end{table}

We use the STEC\textsubscript{LO} model to assess the contribution of intermediate layers to overall performance. Unlike baseline click-through rate (CTR) models, which do not typically shorten the distance between intermediate representations and output, our approach allows lower-order interactions to directly affect prediction output without traversing through higher-order layers. Our results demonstrate that including intermediate bilinear interactions significantly improves performance across all settings.

We also compare various fusion strategies for bilinear interactions, including concatenation and addition. Our results show that concatenating information from varying complexities performs better than addition, although addition (STEC\textsubscript{F}) still outperforms using only the final interaction output (STEC\textsubscript{LO}). This further supports our argument for leveraging intermediate information sources.

Finally, we compare the performance of STEC\textsubscript{NF} and STEC\textsubscript{BL} to evaluate the impact of the FFN inside the attention block on performance. Our results show that for more challenging datasets such as Avazu and Criteo, the FFN can improve performance, while for smaller datasets such as Frappe and MovieLens, it may degrade performance by increasing model complexity.

\subsection{Explainable Recommendations (RQ4)}

Explainability in recommendation algorithms is of paramount importance for several reasons. Firstly, it fosters trust between the user and the system. When users understand why certain recommendations are made, they are more likely to perceive the system as reliable and accurate. Secondly, explainability can enhance user engagement. By providing insights into the reasoning behind recommendations, users may find the system more intriguing and engaging. Therefore, it is imperative to examine how STEC is able to provide transparent explanations to its predictions.

\begin{figure*}[h]
     \centering
     \begin{subfigure}[b]{0.3\textwidth}
         \centering
         \includegraphics[width=\textwidth]{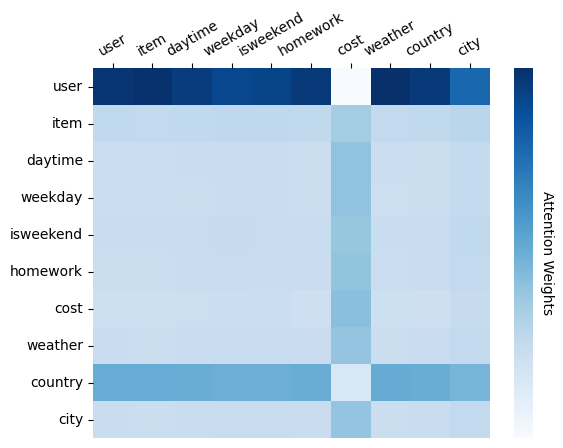}
         \caption{}
         \label{fig:uno}
     \end{subfigure}
     \hfill
     \begin{subfigure}[b]{0.3\textwidth}
         \centering
         \includegraphics[width=\textwidth]{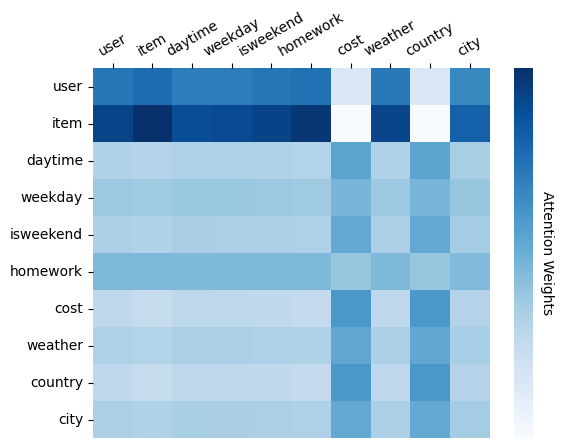}
         \caption{}
         \label{fig:dos}
     \end{subfigure}
     \hfill
     \begin{subfigure}[b]{0.3\textwidth}
         \centering
         \includegraphics[width=\textwidth]{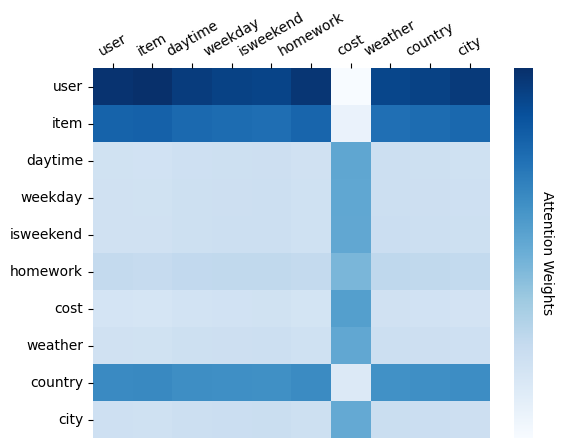}
         \caption{}
         \label{fig:tres}
     \end{subfigure}
        \caption{Heat maps of attention weights for three independent cases on Frappe. The
tick labels correspond to the feature fields \textit{user, item, daytime, weekday, isweekend, homework, cost, weather, country, city}.}
    \label{fig:interactions}
\end{figure*}

To this end, we examine the attention weights of the STEC model on the Frappe dataset. The attention weights for three different scenarios were chosen to demonstrate the expressiveness STEC possesses. Figure \ref{fig:uno} show that the interaction between the user and the other features led the model's prediction. The attention of the STEC model is not limited to only the user representation as seen in Figure \ref{fig:dos} and \ref{fig:tres}. In short, the STEC model is able to leverage arbitrary number of feature interactions to guide its predictions and provide explainable predictions.

\section{Conclusion and Future Work}

We propose a strong and versatile architecture called STEC, which extracts bilinear interaction from attention with minimal overhead to achieve state-of-the-art results. We conducted extensive experiments on four commonly used benchmark datasets and offline, and
online A/B tests.
Our experiments demonstrate that our approach is not dataset-specific and performs well across a range of datasets. Furthermore, we show that introducing direct connections between different interaction levels and output improves performance by reducing the distance between lower-level interactions and output. Additionally, we demonstrate that reformulating existing blocks such as scaled dot-product attention can yield performance gains without incurring additional computational cost.

The significance of explainability in recommendation systems cannot be overstated. It offers a clear view into the system’s operations, helping users understand the rationale behind specific recommendations, which can bolster their confidence in the system. We show that the STEC model can provide explainable and interpretable rationales for its predictions via its attention mechanism all the while staying competitive with models which perform well but provide no means to investigate their inner workings. 

In future work, we plan to investigate whether current stacked architectures of CTR models can be modified to perform higher-order interaction calculations only when lower-order interactions are not sufficient for prediction. Our proposed architecture can potentially accommodate such a structure if the multilayer perceptron (MLP) on bilinear interactions is altered to contain the effect of each interaction within the same layer.

%Bibliography
\bibliographystyle{unsrt}  
\bibliography{aaai24}

\end{document}